\documentclass{aa}
\usepackage{pictexwd,rotating,amssymb,nicefrac}
\usepackage{booktabs,afterpage}
\usepackage[dvips]{color}
\newfont{\tinytiny}{cmr5 at 4pt}
\newfont{\tinytinytiny}{cmr5 at 1pt}

\def\yerrorbar at #1 #2, yerror from #3 to #4 {%
\put {\line(1,0){1}} at #1 #3
\put {\line(1,0){1}} at #1 #4
\plot #1 #3 #1 #4 /
\put {\circle*{0.5}} [Bl] at #1 #2
}
\begin{document}

   \title{A QSO survey via optical variability and zero proper motion
   in the M\,92 field}

   \subtitle{IV.
   More  QSOs due to improved photometry
}
   \author{J. Brunzendorf$\,^{\star}$
          \and
           H. Meusinger
\thanks{
Visiting Astronomer, German-Spanish Astronomical Centre, Calar Alto,
operated by the Max-Planck-Institute for Astronomy, Heidelberg, jointly
with the Spanish National Commission for Astronomy}
}

   \institute{Th\"uringer Landessternwarte Tautenburg, D-07778 Tautenburg\\
              e-mail: brunz@tls-tautenburg.de, meus@tls-tautenburg.de
             }


   \date{}

   \abstract{We continue the QSO search in the 10 square degrees
Schmidt field around M\,92 based on variability and proper motion (VPM)
constraints. We have re-reduced 162 digitised $B$ plates with a 
time-baseline of more than three decades and have considerably improved 
both the photometric accuracy and the star-galaxy separation at $B>19$. 
QSO candidates are selected and marked with one out of three degrees 
of priority based on the statistical significance of their
measured variability and zero proper motion. Spectroscopic follow-up
observations of 84 new candidates with $B>19$ revealed an additional
37 QSOs and 7 Seyfert\,1s. In particular, all 92 high-priority
candidates are spectroscopically classified now; among them are
70 QSOs and 9 Seyfert\,1s (success rate 86\%). We expect that
87\% (55\%) of all QSOs with $B<19.0$ ($19.8$) are contained in this
high-priority subsample. For the combined sample of
high-priority and medium-priority objects, a completeness of 89\% is
estimated up to $B_{\rm lim} = 19.5$.
The sample of all
AGNs detected in the framework of the VPM search in the M\,92 field
contains now 95 QSOs and 14 Seyfert\,1s with $B\le19.9$.
Although the VPM QSOs were selected by
completely different criteria, their properties do not significantly
differ from those of QSOs found by more traditional optical survey
techniques. In particular, the spectra and the optical broad band colours
do not provide any hints on a substantial population of red QSOs up to
the present survey limit.
   }

\authorrunning{Brunzendorf \& Meusinger}
\titlerunning{VPM QSO survey in the M\,92 field}

   \maketitle

%
%
\section{Introduction}
%

The number of known QSOs is growing very rapidly. Over the last decade, 
among others,
the Durham/AAT survey (Boyle et al. \cite{Boy90}), 
the Large Bright Quasar Survey (Hewett et al. \cite{Hew95}),
the Edinburgh Quasar Survey (Goldschmidt \& Miller \cite{Gol98}),
and the Hamburg/ESO survey (Wisotzki et al. \cite{Wis00})
have been completed. Presently, the 2dF Quasar Survey (Croom et al.
\cite{Cro01}) and the Sloan Digital Sky Survey (Schneider et al.
\cite{Sch01}) are extremely efficient at identifying very large
numbers of quasars. The INT Wide Angle
Survey (Sharp et al. \cite{Sha01}) is expected to detect a statistically
significant sample of high-redshift quasars. Very deep quasar samples were
obtained in the Lockman hole via the X-ray satellite ROSAT
(Hasinger et al. \cite{Has98}) and in the optical domain with the
Hubble Space Telescope (e.\,g., Conti et al. \cite{Con98}),
respectively. Further, the VLA FIRST Bright Quasar Survey
(e.\,g., White et al. \cite{Whi00})
will define a radio-selected QSO sample that is competitive in
size with current optically selected samples. All these surveys
select the QSO candidates on the basis of  their particular
colours with respect to stars and galaxies or their brightness at
X-ray or radio wavelengths. In other words, they rely on  the
different broad-band spectral energy distribution (SED) as the prime
selection criterion.

Despite the large number of QSOs now catalogued, the selection effects of the
conventional surveys are not yet fully understood (see, e.g.,
Webster et al. \cite{Web96};
White et al. \cite{Whi00};
Gregg et al. \cite{Gre02}).
It is therefore important to perform alternative QSO surveys
using selection criteria that do not directly rely
on the shape of the SED. Variability and zero proper motion
are such criteria. We started a variabillty and proper motion
(VPM) survey based on several hundred digitised Schmidt plates
(Meusinger et al. \cite{Meu02}). Due to the special demands on
the number and the time-baseline of the available observations
such an attempt must be limited to comparatively small and confined
areas.

We perform the VPM search in two fields of a size of ten square
degrees each, centered on the globular clusters M~3 and M~92,
respectively. The work in the M~92 field is the subject of the
present series of papers. In Paper~1 (Brunzendorf \& Meusinger
\cite{Bru01}), we described the observational material, the
photometric and astrometric data reduction, and the selection
procedure of the QSO candidates. The QSO sample resulting from
the spectroscopic follow-up observations of these candidates
was presented in Paper~2 (Meusinger \& Brunzendorf \cite{Meu01}). In
Paper~3 (Meusinger \& Brunzendorf \cite{MeuB02}), the properties
of the narrow-emission line galaxies among the VPM QSO candidates
are discussed.

Both the completeness and the efficiency of the VPM search
primarily depend on the photometric accuracy (see Paper~1). From
the comparison with other optical QSO samples we find that the
previous VPM QSO sample is virtually complete for brighter
magnitudes, but its completeness drops rapidly at the faint end
($B>19$). In the context of the work on Paper~3, we have refined
substantial parts of the reduction of all digitised plates with
the result of a significantly improved sample of QSO candidates
at fainter magnitudes. In the present paper, we present the
results of the follow-up spectroscopy and
discuss the properties of the
enlarged VPM QSO sample. 

In Sect.~2, we briefly describe the
major revisions of the data reduction procedure. The new candidate
selection and the selection effects are the subjects of
Sect.~3. The new spectroscopic follow-up observations are
described in Sect.~4. Section~5 deals with the newly detected
QSOs and the properties of the enlarged QSO sample. Finally,
conclusions are given in Sect.~6. As in the previous papers, we
adopt $H_0 = 50$\,km\,s$^{-1}$\,Mpc$^{-1}$ and $q_0=0$.

%
\section{Revised data reduction}
%

\begin{table*}
\caption{\label{selection}Selection criteria and number of QSO candidates for
the medium- and high-priority, respectively, subsamples.}
\begin{tabular}{lll}
\toprule
criterion & high-priority & medium-priority \\
          & QSO candidates & QSO candidates \\
\midrule
 %
 %
proper motion         & $I_\mu\le3$      & $I_\mu\le4.3$      \\
overall variability   & $I_\sigma\ge2$   & $I_\sigma\ge1.645$ \\
long-term variability & $I_\Delta\ge2$   & $I_\Delta\ge1.645$ \\
mean $B$ magnitude    & $18\le B\le19.8$ & $18\le B\le19.8$   \\
remark                &                  & without high-priority objects\\
\midrule
number of candidates  & 92               & 53  \\
classified in Paper~2 & 52 (49~QSOs+Seyfert\,1, 3~stars) & 5 (3~QSOs+Seyfert\,1, 1~NELG, 1~star) \\
remaining             & 40               & 48  \\
\bottomrule
\end{tabular}
\end{table*}

In the present paper, the same observational material is used as
in Papers~1 and 2, namely a selection of 208 photographic plates taken
with the Tautenburg Schmidt telescope in the years $1963 - 1997$ and 
digitised by means of the Tautenburg Plate Scanner. The plates, 
the digitisation, and the reduction
procedures have been presented at length in Paper~1. Here we
describe the revised photometric reduction of the 162 plates taken
in the Johnson $B$ system. There are two major revisions made: we
apply a more powerful software package both for the object search and
for the determination of the object parameters, and we average over
the measurements from plates of nearly the same epoch. The
astrometric data, on the other hand, are taken from the previous
study.

Motivated by the results of extensive tests we decided to replace
our previously used reduction software package by the SExtractor
package (Bertin \& Arnouts \cite{Ber96}). These tests have clearly
shown that the SExtractor detects faint objects, on our plates,
with a much higher reliability. In other words, it  reaches a
higher completeness at faint magnitudes with less false detections
caused by the grain noise of the photographic emulsion. In
addition, its photometric (but not its astrometric) accuracy
proved to be better at faint magnitudes.

\begin{figure}[hbpt]
\includegraphics{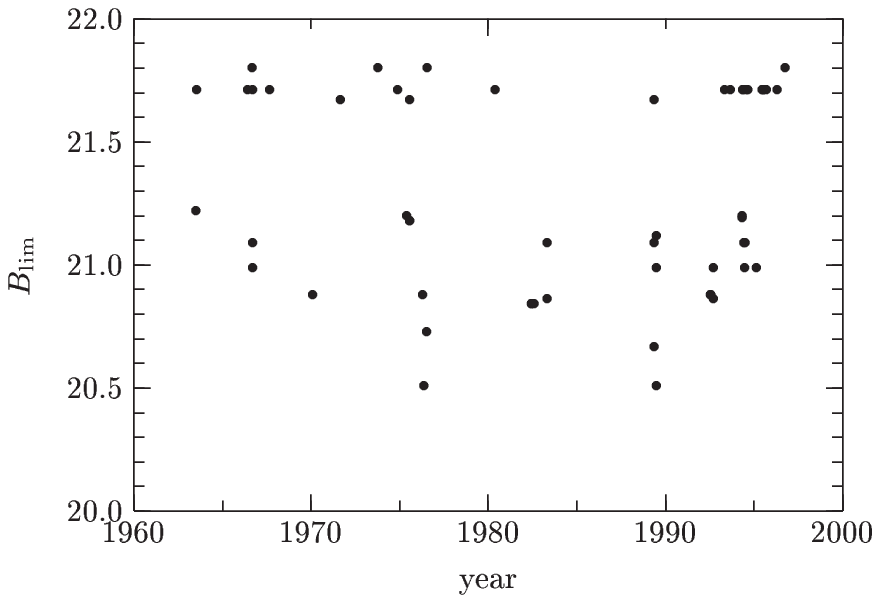}
\caption{\label{epochs}
Limiting magnitudes $B_{\rm lim}$ of the re-reduced photometric data of all
54 epochs.
} 
\end{figure}

The digitisation yields the photographic transparencies $t$ of the
respective plate which have to be transformed into photometric
densities $d=-\log_{10}t$ to provide suitable input data for the
object search realized by the SExtractor. For each plate, in a
first run the full width at half maximum ($FWHM$) of the star-like
sources is derived. The object detection is then done on the image
convolved with a Gaussian of the same $FWHM$ to maximize
the detection sensitivity (see the documentation of the SExtractor
package for details). The final determination of the object
parameters (position, magnitude aso.), however, is done on the
original, i.\,e.  unfiltered, image. SExtractor typically detects
about 30\% more real sources than our previously used software.
The resulting basic object sample, defined in exactly the same way
as described in Paper~1, therefore contains now more than 42\,200
objects.
The additional objects have faint magnitudes down to $B=21$.
For the present, however, we are primarily interested in
a high completeness of the total VPM QSO sample. The main aim
of the present study is to complete the QSO sample in the M\,92
field to the limiting magnitude $B_{\rm lim} \approx 19.7$ from
Papers~1 and 2. For this magnitude
range, the new sample of objects is essentially identical with
the original sample from Paper~1.
The discussion of the selection effects from Paper~1
remains largely valid; modifications due to the improved
photometric accuracy are discussed in Sect.\,3.2. The investigation
of the additional objects has to be deferred to a prospective study.

\begin{figure}[hbpt]
\resizebox{8.8cm}{6.5cm}{\includegraphics{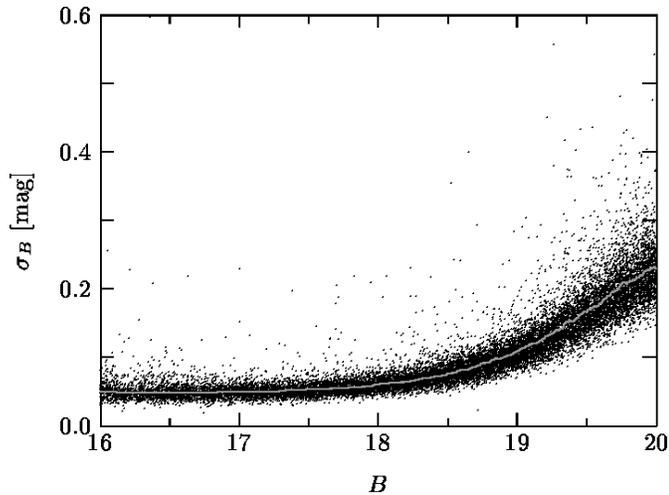}}
\caption{\label{B_mittel_sigmaB} 
Photometric standard deviation $\sigma_B$ of the magnitudes $B_i$ measured for
each star-like object on the plates $i$ as a function of the mean magnitude
$B$. The curve represents the median value.
}
\end{figure}

The photometric calibration of the revised internal magnitudes
was done in exactly the same way as outlined in Paper~1. The
revised photometry yields limiting magnitudes of the plates in the
range of $B_{\rm lim}=19 - 22$,  where $B_{\rm lim}$ is
identified with the $B$ magnitude of the faintest standard star
detected. In order to obtain a maximum
sensibility with respect to faint variable objects, it is
advisable to minimize the impact of the different photometric
accuracies and limiting magnitudes of the individual plates caused
by the scatter in their qualities. For this reason, we only
consider plates that match {\em all} of the following criteria:
(1.) limiting magnitude $B_{\rm lim}\ge19.8$, (2.) number of
detected objects $N\ge30\,000$, and (3.) plates taken at zenith
distances $z\le45\degr$. These criteria are satisfied by 116 $B$
plates. We aim at a photometric accuracy of better than
$\sim0.1$\,mag at $B=19$ and better than $\sim0.2$\,mag at $B=20$.
In order to reach this aim, we averaged 18 pairs of plates of
close-by epochs as well as 10 triples of plates of adjacent
epochs. In addition, we excluded further 26 plates from our data
sample, since in these cases plates of the same or close-by epochs
yet significantly better quality are available. The final
photometric data set comprises 54 epochs spanning 33.2 years,
where each single plate
matches the quality constraints defined above. An
overview over the final epochs and limiting magnitudes $B_{\rm
lim}$ is given in Fig.~\ref{epochs}. The mean accuracy of the 
photometric measurements of each object is shown in 
Fig.~\ref{B_mittel_sigmaB}. The comparison
with Fig.~5 in Paper\,1 indicates a substantial reduction of the
photometric errors, especially for $B>19$. The calculation of
the new indices for overall variability $I_\sigma^{\rm new}$ and
long-term variability $I_\Delta^{\rm new}$ strictly follows the
descriptions given in Paper~1. In the following, we refer to these
new variability indices as $I_\sigma$ and $I_\Delta$ without the
index $^{\rm new}$. Note however, that the values of these indices
are not identical with the data given in Papers~1 and 2.

\begin{figure}[hbpt]
\resizebox{8.8cm}{6.5cm}{\includegraphics{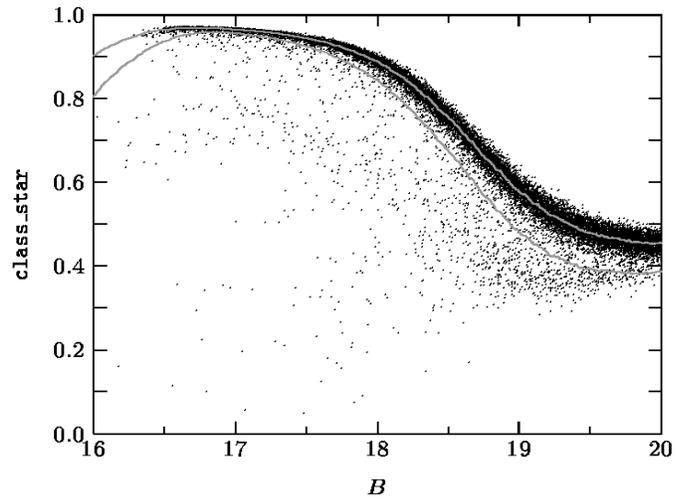}}
\caption{\label{class} 
Classification parameter {\tt class\_star} for all measured objects in the
magnitude range $B=16-20$. The curves represent the median value (upper) 
and the 5$\sigma$ deviation from the median (lower).
}
\end{figure}

An effective star-galaxy separation is a further important
requirement  for an efficient candidate selection, mainly because
deviations from  the star-like appearance may produce additional
photometric errors that can  mock variability (Paper~3).
In our first attempt (Paper~1), image profile indices were derived
on the basis of the relation between radii and magnitudes
measured on a deep $R$ plate. The SExtractor package allows a
more sophisticated morphological classification based on a trained
neural network.  Moreover, in Paper~3, we have stressed that the
classification should be done on exactly the same plates which
are used for photometry and/or astrometry.
In the present study, we use the index {\tt class\_star}
computed by SExtractor from the measurements on the $B$ plates.
An object is considered stellar if ${\tt class\_star} < 0.4$,
which roughly corresponds to a $5\sigma$ deviation from
the median value in the relevant magnitude range $19 \le B \le 19.8$
(Fig.\,\ref{class}).  In this way, the star-galaxy separation
is substantially improved.

%
\section{QSO candidate selection and selection effects}
%


\subsection{QSO candidate selection}


\begin{figure}[hbpt]
\resizebox{8.0cm}{5.5cm}{\includegraphics{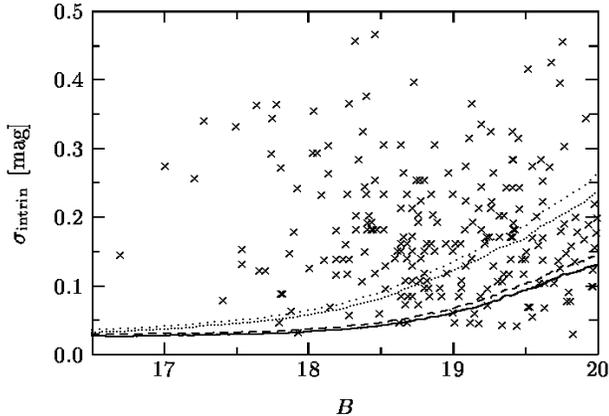}}
\caption{\label{Hook_detectionlimit}
Standard deviation of the intrinsic magnitude fluctuation, 
$\sigma_{\rm intrin}$,  versus apparent mean $B$ magnitude 
for the QSOs ($\times$) from Hook et al. (\cite{Hoo94}). 
The curves indicate the detection limits $\sigma_{\rm intrin}^{\rm min}$ of the 
present paper for $I_\sigma\ge2$ (dashed) and $I_\sigma\ge1.645$ (solid), 
respectively. For comparison, detection limits from Paper~1 are given (dotted
lines). 
}
\end{figure}

As in Paper~1, an object is considered a QSO candidate if it
shows a star-like image structure (see above) and a
proper motion index $I_\mu\le4.3$. Objects in the dense cluster region
(distance to the centre of M\,92 $r\le6'$) or near the border of the
field (distance to the border $r\le10\,$mm) 
as well as the brightest ($B<13$) and faintest ($B>19.8$) 
objects were excluded.
The remaining 7\,090 of all 34\,500 objects form the
QSO candidate sample. It is expected that {\em all} QSOs
with $13<B<19.8$ in the field are recorded in this sample.
Note that till now no variability constraint is applied.

Among all QSO candidates, a high-priority subsample as well as a
medium-priority subsample are selected according to the constraints given in
Table~\ref{selection}. 
The remaining QSO candidates are of only low priority, since they
have no significant variability. Yet they are
of special interest in conjunction with other QSO selection criteria,
e.\,g., based on colour, when one focuses on the completion of the sample 
of known QSOs in the field.
The magnitude limit $18\le B\le19.8$ 
for the high- and medium-priority candidates 
is motivated by the
fact that, in this paper, we aim at an increase of the completeness at faint
magnitudes. In Papers~1 and 2, we have selected and classified high- and
medium-priority candidates with $B\ge16.5$. In the context of the evaluation
of the selection effects in Paper~1 as well as by a direct comparison with 
the mean QSO surface density from other surveys (Paper~2), we have shown 
that our QSO sample is virtually complete, at least for $B\le18$. 
In the present paper, we consider
therefore only candidates with $B\ge18$. The slight increase in the magnitude limit
$B_{\rm lim}$ from 19.7 (Paper~1) to 19.8 is justified by the
better photometric accuracy at faint magnitudes.
Among the remaining low-priority candidates, a number of objects might be of
special interest as well. These objects do either meet the classical colour
selection criteria, e.\,g. show a UV excess, and/or failed only one of the
variability criteria.

\begin{figure}[hbpt]
\resizebox{8.6cm}{6.1cm}{\includegraphics{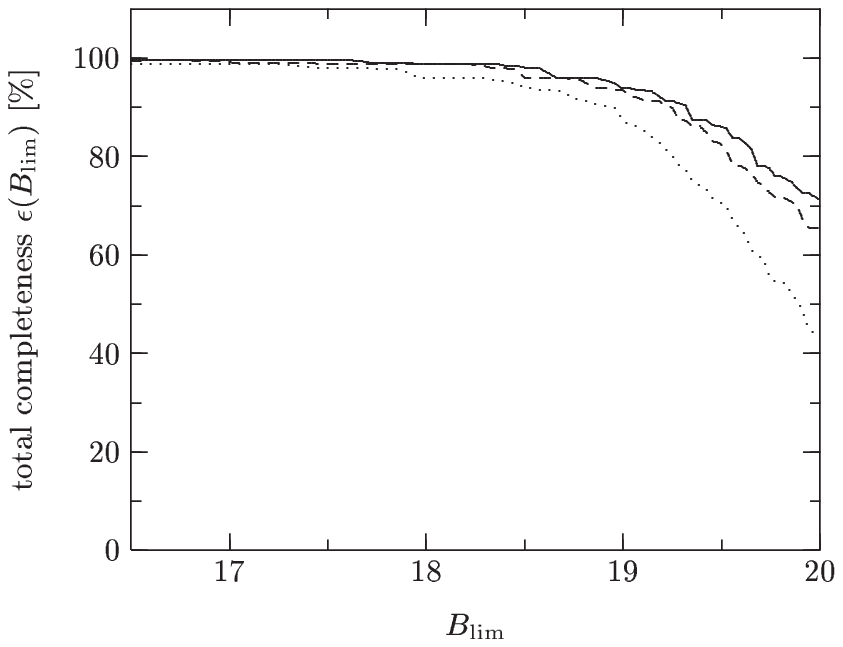}}
\caption{\label{completeness}
Expected total completeness $\epsilon(B_{\rm lim})$ of the present survey 
up to a given limiting magnitude $B_{\rm lim}$. It is assumed 
that the QSO candidate selection is based on the $I_{\sigma}$ criterion 
only with $I_\sigma\ge2$ (dashed curve) and $I_\sigma\ge1.645$
(solid curve), respectively. The dotted indicates the estimated completeness 
for the selection criterion $I_\Delta\ge2$ derived from a simple model assuming 
sinusoidal brightness fluctuations with periodicies $T\approx3$\,yr. 
In that case, all objects with periodicities $T\lesssim30$\,yr also meet the 
criterion $I_\sigma\ge2$.
}
\end{figure}

\begin{figure}[hbpt]
\resizebox{8.4cm}{5.8cm}{\includegraphics{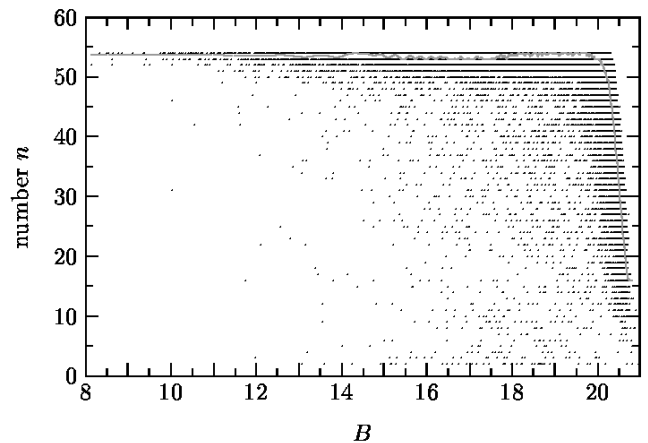}}
\caption{\label{anzahl_epochen}
Number of observations $n$ for each object as a function of its mean apparent
magnitude $B$. The tight correlation is illustrated by the median curve (solid).
} 
\end{figure}

\begin{table*}
{\renewcommand{\baselinestretch}{0.95}\footnotesize
\caption{\label{QSO_list}
Additional QSOs and Seyfert\,1 galaxies from the present study. 
In the columns on the right hand side, the index defined in paper\,1 for 
proper motion ($I_\mu$) as well as the newly calaculated overall variability 
index $I_\sigma$ and long-term variability index $I_\Delta$ are given.
}
\begin{tabular}{rllrrrrrrrr}
\toprule
no & $\alpha$\,(J2000) & $\delta$\,(J2000)  & \multicolumn{1}{c}{$z$} & $M_B$ & \multicolumn{1}{c}{$B$} 
& $\overline{U-B}$ & $\overline{B-V}$ & \multicolumn{1}{c}{$I_\mu$} & \multicolumn{1}{c}{$I_\sigma$} 
& \multicolumn{1}{c}{$I_\Delta$} \\
\midrule
QSO\\
\midrule
  66 & 17 08 46.18 & 44 08 39.7  &   1.544  &   --25.8 &   19.6 & --0.66 &   0.52 &  1.19 &   9.05 &  3.14 \\ 
  67 & 17 09 42.45 & 42 53 14.4  &   0.705  &   --23.8 &   19.7 & --0.49 &   0.10 &  1.23 &   4.53 &  1.45 \\ 
  68 & 17 10 01.16 & 43 25 48.2  &   2.155  &   --26.7 &   19.9 & --0.82 &   0.04 &  2.06 &   6.25 &  3.80 \\ 
  69 & 17 10 52.64 & 43 15 05.9  &   0.749  &   --24.0 &   19.7 & --1.19 &   0.39 &  3.09 &   2.08 &  2.34 \\ 
  70 & 17 10 55.63 & 42 53 09.5  &   1.088  &   --24.9 &   19.8 & --0.87 &   0.04 &  2.46 &   4.04 &  3.57 \\ 
  71 & 17 11 23.89 & 43 48 37.7  &   1.838  &   --26.7 &   19.1 & --0.50 &   0.06 &  2.77 &   0.78 &  2.08 \\ 
  72 & 17 12 00.57 & 43 29 44.2  &   1.457  &   --25.9 &   19.3 & --1.13 &   0.17 &  1.91 &   5.15 &  2.20 \\ 
  73 & 17 12 26.39 & 41 42 12.0  &   1.148  &   --25.1 &   19.8 & --0.76 & --0.32 &  0.43 &   4.95 &  2.88 \\ 
  74 & 17 13 08.81 & 42 08 11.4  &   2.300  &   --27.0 &   19.6 & --0.58 &   0.51 &  1.91 &   4.07 &  4.95 \\ 
  75 & 17 13 10.63 & 43 48 55.6  &   2.484  &   --27.1 &   19.7 & --0.53 &   0.47 &  2.15 &   2.38 &  3.54 \\ 
  76 & 17 13 11.19 & 42 51 54.2  &   0.959  &   --24.7 &   19.7 & --0.75 &   0.12 &  0.89 &   3.89 &  3.82 \\ 
  77 & 17 13 20.88 & 44 15 31.1  &   0.887  &   --24.7 &   19.3 & --1.06 &   0.35 &  2.86 &   5.83 &  1.90 \\ 
  78 & 17 14 02.85 & 42 44 22.8  &   1.494  &   --25.8 &   19.5 & --0.82 &   0.05 &  1.90 &   6.11 &  2.79 \\ 
  79 & 17 14 04.14 & 43 22 36.4  &   1.565  &   --25.8 &   19.8 & --0.76 &   0.16 &  1.27 &   7.93 &  4.88 \\ 
  80 & 17 15 00.59 & 42 54 39.4  &   1.508  &   --26.0 &   19.5 & --0.52 & --0.15 &  1.35 &   4.41 &  4.24 \\ 
  81 & 17 15 32.18 & 43 47 59.4  &   0.923  &   --24.6 &   19.5 & --0.83 &   0.27 &  1.64 &   3.28 &  3.63 \\ 
  82 & 17 15 40.06 & 42 42 26.8  &   1.835  &   --26.3 &   19.7 & --0.97 & --0.02 &  0.30 & --1.51 &  1.82 \\ 
  83 & 17 16 05.82 & 43 30 26.8  &   2.898  &   --27.9 &   19.8 & --0.40 &   0.39 &  0.24 &   1.73 &  3.14 \\ 
  84 & 17 17 13.20 & 44 21 04.7  &   0.607  &   --23.2 &   19.8 & --1.15 &   0.43 &  0.75 &   4.78 &  3.08 \\ 
  85 & 17 17 32.55 & 43 56 55.6  &   0.677  &   --24.0 &   19.3 & --0.98 &   0.34 &  0.84 &   4.73 &  2.97 \\ 
  86 & 17 17 36.75 & 44 23 26.9  &   1.606  &   --25.9 &   19.7 & --1.09 &   0.47 &  0.88 &   6.23 &  4.21 \\ 
  87 & 17 17 36.95 & 42 16 53.8  &   1.847  &   --26.4 &   19.6 & --1.22 &   0.57 &  1.30 &   2.49 &  3.42 \\ 
  88 & 17 18 24.00 & 42 25 47.7  &   1.688  &   --26.2 &   19.5 & --0.82 &   0.38 &  2.28 &   1.12 &  2.75 \\ 
  89 & 17 19 12.04 & 44 11 22.1  &   1.782  &   --26.5 &   19.4 & --0.75 & --0.33 &  0.87 &   2.12 &  2.42 \\ 
  90 & 17 19 12.23 & 43 20 44.7  &   2.603  &   --28.1 &   18.9 & --0.18 & --0.14 &  1.12 &   6.63 &  3.43 \\ 
  91 & 17 19 15.42 & 44 22 46.9  &   0.420  &   --23.4 &   18.6 & --0.69 &   0.30 &  1.94 &  12.08 &  2.02 \\ 
  92 & 17 19 42.87 & 43 09 27.1  &   3.199  &   --28.7 &   19.5 &   1.51 &   1.11 &  0.77 &   7.87 &  4.28 \\ 
  93 & 17 19 52.94 & 41 43 39.5  &   1.484  &   --25.7 &   19.8 & --0.67 &   0.63 &  0.49 &   6.03 &  2.88 \\ 
  94 & 17 20 04.81 & 41 47 44.1  &   1.218  &   --25.1 &   19.8 & --0.90 &   0.36 &  3.16 &   6.66 &  3.24 \\ 
  95 & 17 20 11.02 & 43 46 34.9  &   1.459  &   --25.5 &   19.9 & --1.59 &   0.56 &  1.96 &   4.45 &  4.09 \\ 
  96 & 17 21 17.57 & 43 01 27.0  &   0.526  &   --23.2 &   19.5 & --0.75 &   0.26 &  0.95 &   0.05 &  1.81 \\ 
  97 & 17 21 42.18 & 42 44 32.5  &   2.306  &   --26.9 &   19.4 & --0.53 &   0.11 &  2.08 &   4.12 &  2.11 \\ 
  98 & 17 22 03.06 & 43 41 06.3  &   1.069  &   --25.1 &   19.4 & --1.00 &   0.14 &  2.57 &   6.83 &  1.19 \\ 
  99 & 17 23 21.49 & 43 29 03.3  &   2.030  &   --26.5 &   19.9 & --0.69 & --0.31 &  2.87 &   6.61 &  3.41 \\ 
 100 & 17 23 45.96 & 44 03 27.2  &   1.749  &   --26.0 &   19.7 & --0.87 &     -- &  1.38 &   5.40 &  3.93 \\ 
 101 & 17 24 01.51 & 43 20 30.5  &   1.463  &   --25.7 &   19.7 & --0.83 &   0.50 &  1.01 &   3.14 &  4.28 \\ 
 102 & 17 24 14.22 & 42 10 47.2  &   0.748  &   --24.0 &   19.5 & --1.07 &   0.42 &  0.38 &   6.20 &  3.03 \\ 
\midrule
Sy\,1 \\
\midrule
 103 & 17 11 39.81 & 42 14 51.5  &   0.175  &   --20.9 &   19.5 & --0.26 &   1.43 &  1.78 &   8.45 &  3.74 \\ 
 104 & 17 14 45.16 & 42 20 27.4  &   0.337  &   --22.2 &   19.5 & --1.09 &   0.27 &  1.40 &   7.48 &  2.59 \\ 
 105 & 17 16 06.82 & 44 08 35.9  &   0.439  &   --22.8 &   19.4 & --0.93 & --0.27 &  1.62 &   6.54 &  3.70 \\ 
 106 & 17 18 20.94 & 42 49 13.6  &   0.166  &   --21.3 &   18.7 & --0.60 &   0.41 &  3.44 &  24.71 &  4.67 \\ 
 107 & 17 19 11.96 & 42 10 48.0  &   0.169  &   --20.6 &   19.7 & --0.34 &   1.26 &  2.48 &   9.14 &  3.11 \\ 
 108 & 17 19 36.63 & 42 45 29.1  &   0.284  &   --21.9 &   19.5 & --0.89 &   0.83 &  0.85 &   3.33 &  3.81 \\ 
 109 & 17 21 53.87 & 42 33 36.3  &   0.209  &   --21.7 &   18.9 & --0.75 &   0.72 &  2.15 &   8.96 &  2.59 \\ 
\bottomrule                                         
\end{tabular}
}
\end{table*}

\begin{figure*}[hbpt]
\resizebox{16.3cm}{20.1cm}{\includegraphics{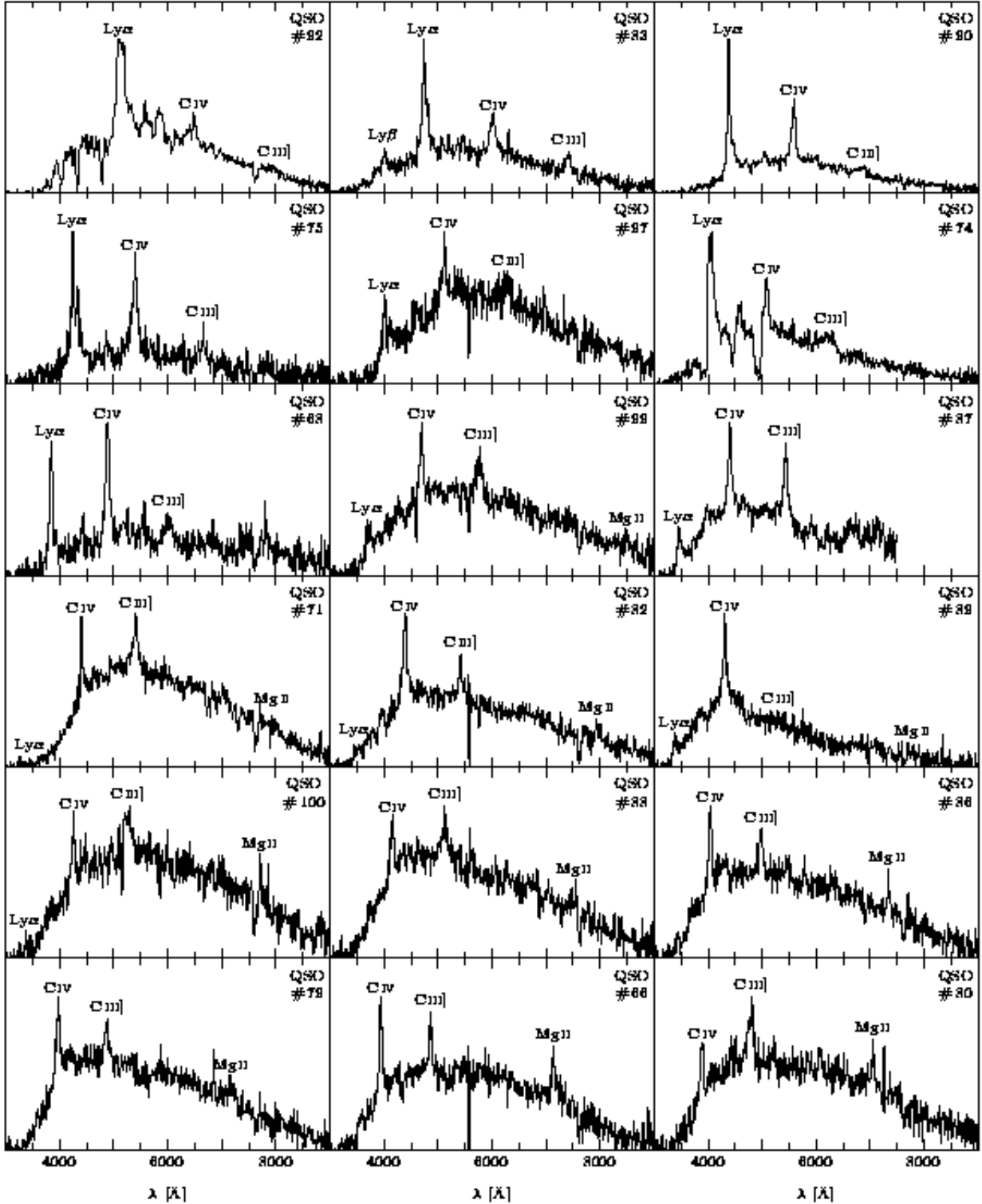}}
\caption{\label{allspectra}
Low-resolution spectra (uncalibrated flux $f_{\lambda}$ versus wavelength
$\lambda$) of all
QSOs and Seyfert\,1s from the present study. In the upper right corner
of each panel, the object type and the running number from 
Tab.~\ref{QSO_list} are given.
}
\end{figure*}
 
\begin{figure*}[hbpt]
\resizebox{16.3cm}{22.6cm}{\includegraphics{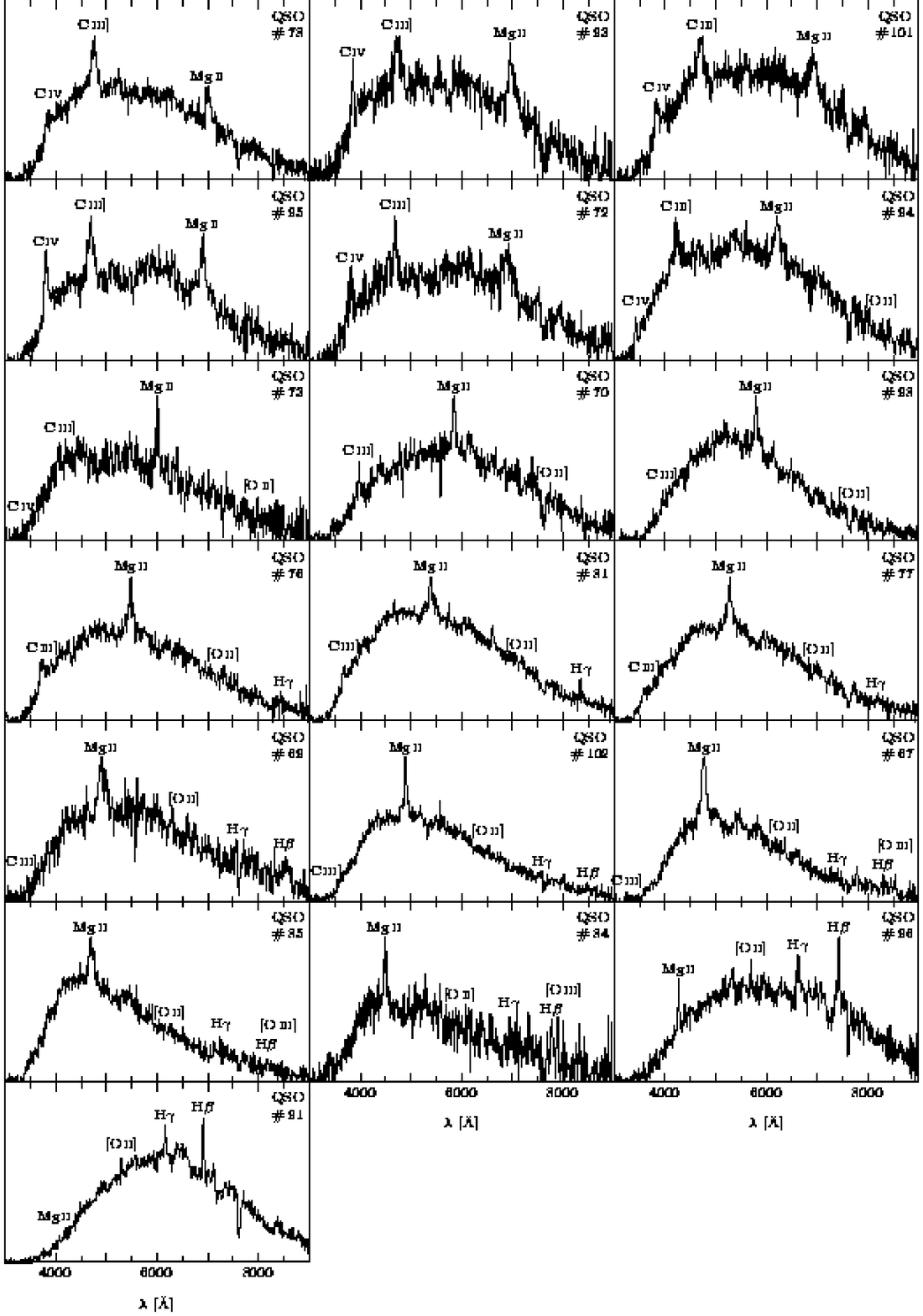}}
\addtocounter{figure}{-1}
\caption{(continued)
}
\end{figure*}

\begin{figure*}[hbpt]
\resizebox{16.3cm}{10.5cm}{\includegraphics{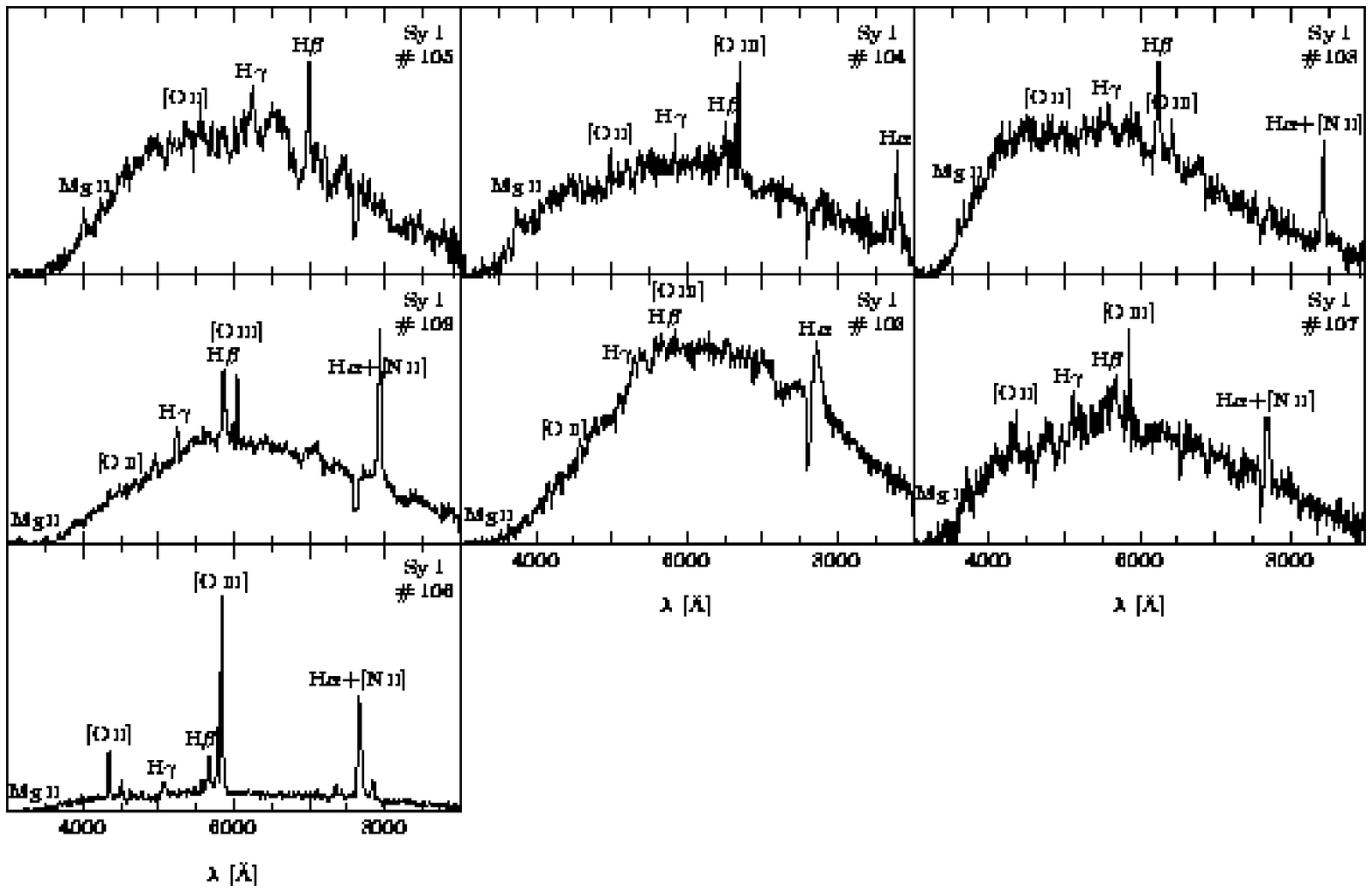}}
\addtocounter{figure}{-1}
\caption{(continued)
}
\end{figure*}

%
\subsection{\label{sect_selection_effects}Selection effects}
%

With respect to the better photometric accuracy and the reduced number 
of epochs we repeat here the estimation of the completeness rates outlined
in Paper~1. In Fig.\,\ref{Hook_detectionlimit}, we compare
the intrinsic magnitude fluctuations measured
by Hook et al. (\cite{Hoo94}) for a large samples of conventionally selected 
QSOs with the
detection limits of the present study. For now we only consider the
incompleteness caused by the application of the overall variability criterion
$I_\sigma>2$, or $I_\sigma>1.645$, respectively.
Considering the Figs.\,20 and 21 in Paper~1, which refer to the variability
properties of the Hook-QSOs, Eq.\,(23) in Paper~1, and the photometric
accuracy of our revised data (Fig.\,\ref{B_mittel_sigmaB}), we derived
the revised completeness rates for the selection based on $I_\sigma$.
The results are shown in Fig.\,\ref{completeness}. 
A direct comparison with
Fig.\,19 in Paper~1 clearly reveals an improved completeness at faint
magnitudes. Now, more than 74\% of all QSOs brighter than $B_{\rm lim}=19.7$
are expected to be included in the high-priority candidate subsample
($I_\sigma>2$), compared to only 64\% in Paper~1. Due to the better
photometric accuracy, the expected completeness at 
$B_{\rm lim}=20$ amounts to 65\%, compared with previously 52\% 
(Paper~1).

The number of epochs is, of course, significantly smaller than in Paper~1
(54 {\em versus} 152). Nevertheless, this reduction is not expected to
provide a substantial selection effect: for objects brighter than  
$B = 20$, the number of photometric data points is sufficiently constant
(Fig.~\ref{anzahl_epochen}). Thus,  the impact of the varying number of 
epochs on the total selection effects is neglibible for $B\le20$. Note that
at $B\approx20$ the mean number of data points is about 50 in this paper
as well as in Paper~1, yet the photometric accuracy is
significantly higher in this paper.

The investigation of the selection effects caused by the application of the
long-term variability index $I_\Delta$ is not as simple as for the overall
variability index $I_\sigma$, since it depends not only on the amplitude of
the intrinsic variability of the object in units of the photometric accuracy
at this magnitude, but also on the shape of the lightcurve, in
particular on the dominating time-scale of the variability. We applied
similar simple model simulations as described in Paper~1 to illustrate 
the influence of $I_\Delta$. Lightcurves
are calculated for the 54 epochs of the present study assuming a
sinusoidal process with periodicity $T$ superimposed on white noise.
The variability indices $I_\sigma$ and $I_\Delta$ directly result
from the chosen values of $T$ and the amplitude of the sinusoidal
in units of the noise. If $T\lesssim30$\,yr, the selection of the
QSO candidates is defined by the constraint $I_\Delta\ge2$,  since in
this case the overall variability is always $I_\sigma>2$. In order to
provide at least an estimate for the incompleteness caused by the selection
criterion $I_\Delta\ge2$ we assumed a typical time-scale of
$T\approx$3\,yr (observer frame). The resulting completeness rates
are shown in Fig.\,\ref{completeness}.

%
\section{Spectroscopic follow-up observations}
%

\begin{figure*}[hbpt]
\resizebox{17.8cm}{6cm}{\includegraphics{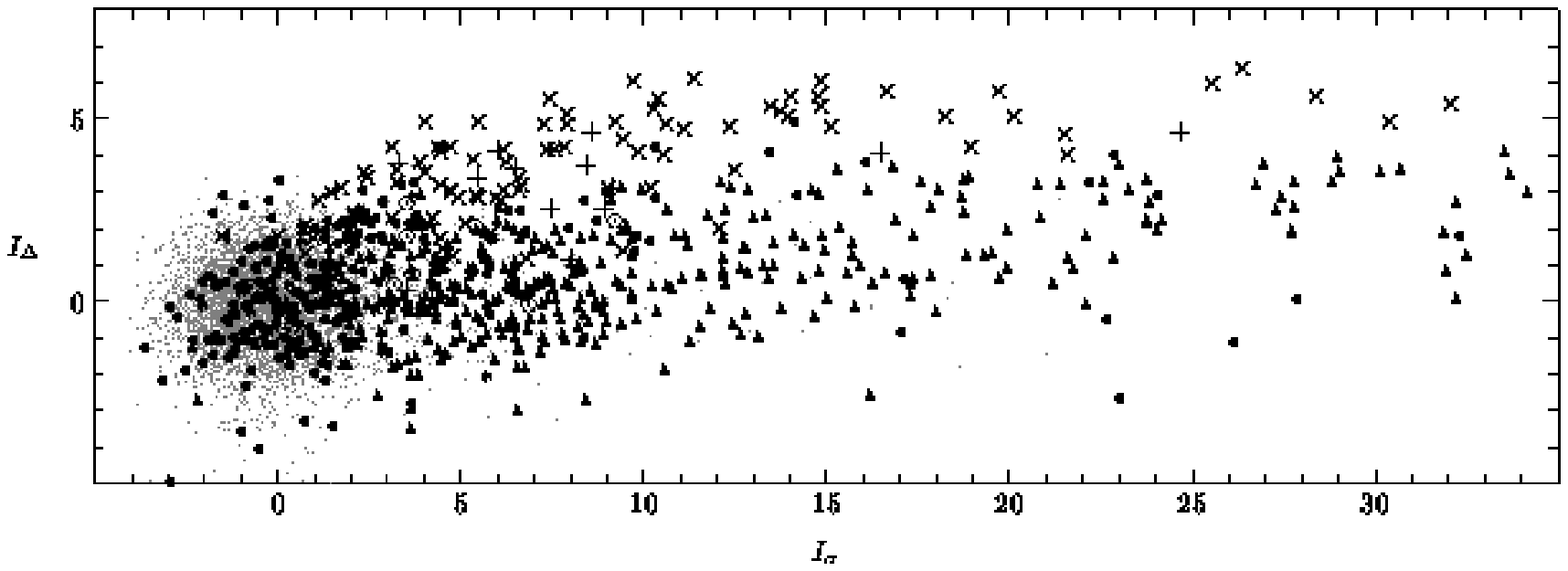}}
\caption{\label{Isigma_Idelta}
Long-term variability index $I_\Delta$ versus overall 
variability index $I_\sigma$ for all QSO candidates 
with $B\ge16.5$ (gray dots), established QSOs ({\normalsize$\times$}),
Seyferts ({\normalsize$+$}), NELGs ({\small$\ast$}), 
spectroscopically identified stars ({\tiny$\bullet$}), 
visually identified galaxies ({\tiny$\blacktriangle$}), 
and visual double stars ({\scriptsize$\sun$}).
}
\end{figure*}

Spectroscopic follow-up observations of the new QSO candidates
were performed again with CAFOS at the 2.2~m telescope on Calar Alto,
Spain.  During six nights in July, 2001, low-resolution spectra were 
taken with the B-400 grism and a SITe1d CCD. The weather conditions 
were good, partly photometric,  with the exception of a cloudy last 
night. The strategy of the observations closely followed
the procedures of the 1998 and 1999 campaigns (Paper~2).
Depending on the seeing of $1\arcsec\ldots2\arcsec$,
a slit width of $1\farcs2\ldots1\farcs7$ was chosen, equivalent to an
effective linear resolution of 22\ldots31\AA{}. At the beginning of each night,
calibration lamp spectra were taken to provide information for the data
reduction, in particular for the wavelength calibration.

\noindent
In total, 84 new QSO candidates of the M\,92 field were successfully observed,
namely: 

\vspace{-0.2cm}
\begin{itemize}
\item[({\bf a})] all 40 remaining high-priority candidates,
\item[({\bf b})] 9 further candidates that match all requirements for the 
      high-priority subsample except that they are slightly fainter 
      than $B=19.8$,
\item[({\bf c})] 23 of the 48 medium-priority candidates,
      among them all objects  fainter than $B=19$ as well as all objects
      with $U-B\le-0.4$,
\item[({\bf d})] all 10 low-priority candidates fainter than $B=19.0$
      that have at least one variability index, $I_\sigma$ or $I_\Delta$,
      greater than 1.645 and meet the colour selection criterion given by
      Scholz et al. (\cite{Sch97}),
\item[({\bf e})] 2 low-priority candidates that meet the
      colour selection criterion by Scholz et al. (\cite{Sch97}).
\end{itemize}

The reduction of the spectra was done under ESO-MIDAS using the
long-slit spectroscopy package
{\tt LONG}. As in Paper~2, we abstained from a flux calibration.
The classification of all candidates with spectroscopic follow-up
observations was done as described in Paper~2. In one case (J171926.5+432123,
not contained in Table~\ref{QSO_list}) we could not unambiguously classify the
spectrum due to the poor signal-to-noise ratio. With only one emission line,
without clear evidence for a broad line component, it could have been
a QSO at $z=1.6$ or a galaxy
at $z=0.12$. Since the object appeared fuzzy
and slightly elongated on the acquisition image, we classified it as galaxy.
In all other cases the classification is based on 
at least two emission and/or absorption lines in the respective
spectrum. We also cross-checked if the classification is in general
agreement with the visual morphological classification on the acquisition
frames.

%
\section{Properties of the extended QSO sample}
%

The follow-up spectroscopy revealed an additional 37 QSOs and 
7 Seyfert\,1s
among the 84 observed candidates. The redshifts, magnitudes,
colours, and indices are listed in Table\,\ref{QSO_list} 
in a similar style as in Paper~2. A search in the NED has shown
that  the Seyfert\,1 no.\,106 was previously known as
FBQS\,J171820.9+424914 with $z=0.167$ (White et al. \cite{Whi00}). 
No other identifications with objects of known redshift were found.
The QSO no.71 is within less than $10''$ from the radio sources 
B3 1709+438 and  FIRST J171123.2+434844, and the Seyfert\,1 no.108
is within the error ellipse of the ROSAT source 1RXS~1735.9+424518.
At least three further objects (no. 72, 75, 81)
can be identified with ROSAT sources
from the list of Dahlem \& Thiering (\cite{Dah00}).

\begin{figure}[hbpt]
\resizebox{8.8cm}{6.2cm}{\includegraphics{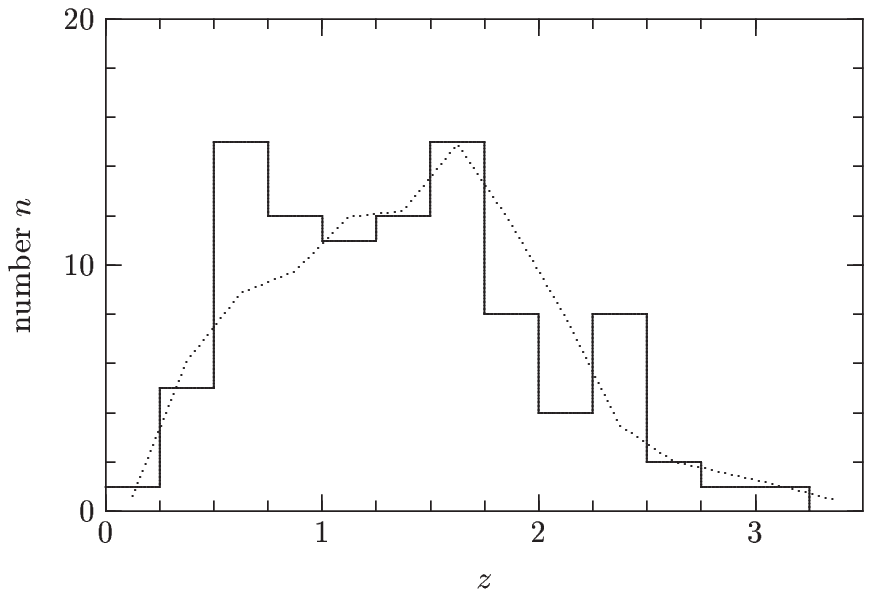}}
\caption{\label{z_hist}
Distribution of the redshifts (number of QSOs per 0.25 redshift bin)
for the extended VPM QSO sample (solid). For comparison, the dotted
polygon gives the $z$ distribution for the 3\,814 QSOs from the early 
data release of the SDSS Quasar Catalogue (Schneider et al. \cite{Sch01}),
normalised to the number of the VPM QSOs. 
} 
\end{figure}

Remarkably, all objects
from Table\,\ref{QSO_list} are fainter than $B=19$, in agreement
with the fact that the major
improvement of the photometry is at the fainter magnitudes
(Sect.~\ref{sect_selection_effects}). With regard to the 
the spectroscopic observations of the different
selection groups listed in the previous Section, the number of newly found
QSOs among the candidates observed is itemised as follows:
({\bf a}) 30 QSOs/Seyfert\,1s out of 40 objects,
({\bf b}) 3 out of 9,
({\bf c}) 5 out of 23, where all five have at least one variability index 
          larger than 2,
({\bf d}) 6 out of 10 , where 4 have one variability index larger than 2, and
({\bf e}) 0 out of 2.

\begin{figure*}[hbpt]
\resizebox{13.2cm}{15.4cm}{\includegraphics{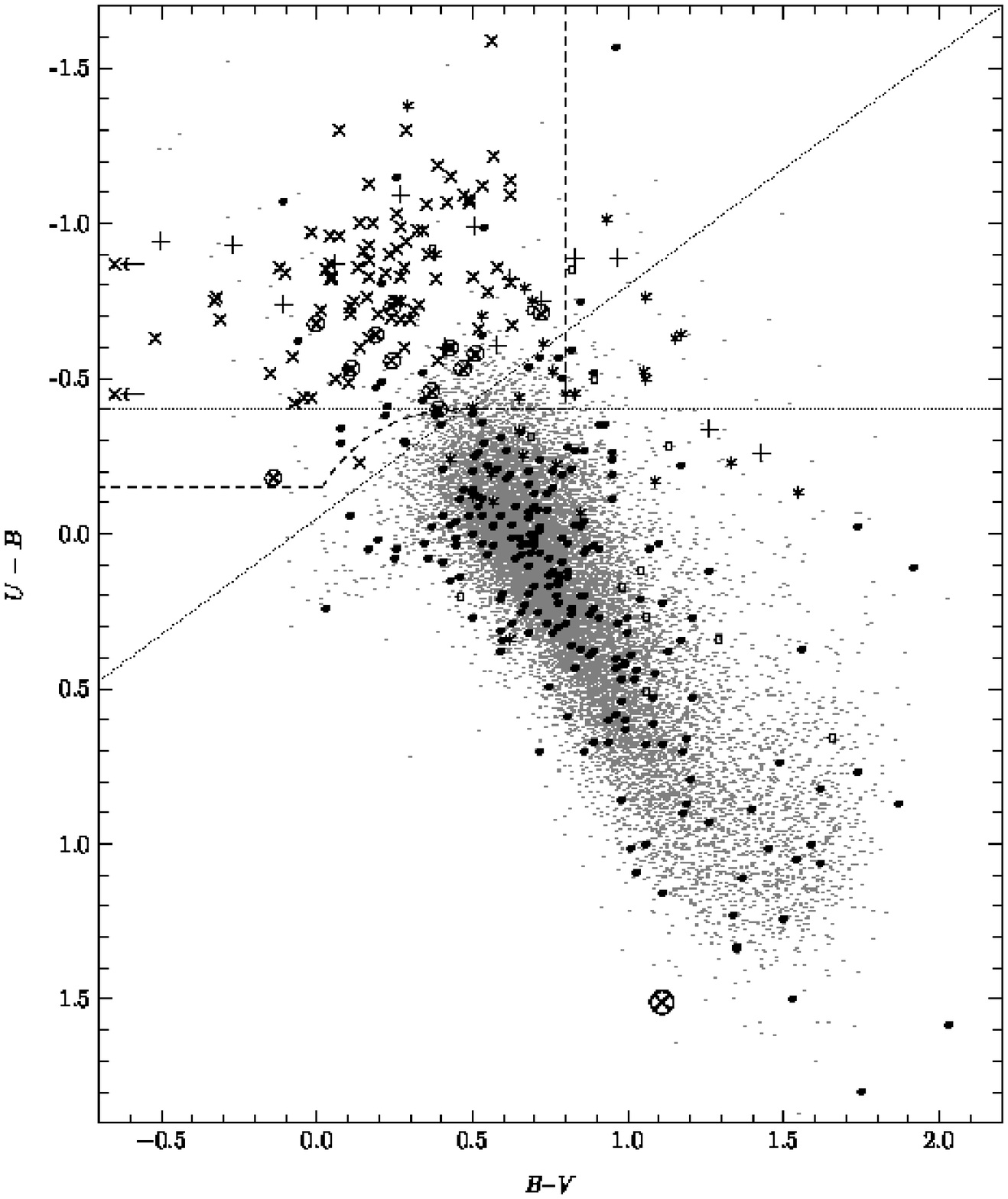}}
\caption{\label{colour_colour}
\color{black}
Colour-colour diagram of the QSOs ($\times$ for $z<2.2$, $\otimes$ for 
$2.2\le z < 3$, $\bigotimes$ for $z > 3$), 
the Seyfert\,1s (+), and the NELGs ($\ast$). 
(Two QSOs without measured $V$ magnitudes are marked by $\leftarrow$ at
the left hand side of the diagram.)
In addition,
spectroscopically identified stars ($\bullet$) and visual binaries
($\circ$) are shown as well as the non-identified objects
with $B \le 19$ (small dots).  
Selection criteria from colour surveys are indicated. 
{\it Horizontal dotted line:} UVX search, 
{\it diagonal dotted line:} 
two-colour search as discussed by Scholz et al. (\cite{Sch97}), 
{\it dashed curve:} two-colour selection according to LaFranca et al.
(\cite{LaF92}).
}
\end{figure*}

With the exception of the high-priority subsample, which was completely
observed, the sample of the further objects observed may appear somewhat 
patchy and inconsequent, in particular because
the VPM search is partly combined with a colour criterion.
The reason for the latter was that it is one of the main intentions
of our quasar search to provide a QSO sample useful for the 
statistical investigation of the long-term variability. For this aim,
it is necessary to avoid a substantial bias against low-variability
QSOs, if they exist. Nevertheless, variability and zero proper motion 
are the dominating constraints for all selection groups but ({\bf e}), which is
however irrelevant. Note also that most (75\%) of the newly detected QSOs/Sy1s 
are found without additional colour constraints and that the statistics 
mentioned above confirms again the efficiency of the variability selection.
An evaluation of the variability indices $I_\sigma$ and $I_\Delta$ of all
objects classified so far (including those from Paper~2) clearly indicates the
distinctive properties of the QSOs/Seyfert\,1s (Fig.~\ref{Isigma_Idelta}) and
confirms again the importance of the long-term variability index for the 
VPM search.

With these new detections, the extended VPM AGN sample in the M\,92 field 
now contains 95 QSOs and 14 Seyfert\,1s. The redshifts of the QSOs span the range
from $z=0.15$ to $z=3.2$, the majority (83\%) have $z<2$. The histogram of the  
redshifts (Fig.\,\ref{z_hist}) is similar to that for the previous sample
(Fig.\,6 in Paper~2), except that the fraction of $z>2$ QSOs is  
larger in the new (22\%) than in the previous data (14\%),
as was expected (Paper~2).  The general form of the $z$ distribution
is comparable with that from the SDSS Quasar Catalogue I. Early 
Data Release (Schneider et al. \cite{Sch01}). The $\chi^2$-test
confirms  at a significance level $\alpha=0.95$
that the two distributions
are not significantly different. In principle, a bias against higher
redshifts is expected for the VPM sample due to the  empirical 
anti-correlation between variability and luminosity/redshift
(e.\,g. Hook et al. \cite{Hoo94}).
Figure\,\ref{z_hist} suggests that such a bias is probably not
strong in our sample. Both for the VPM sample and the SDSS sample,
the distribution has a maximum at $z \approx 1.6$ and decreases
sharply towards higher redshifts. The local maximum at $z\approx 0.6$
in the VPM sample is probably related to the increase of the
survey depth for those $z$ where the strong \ion{Mg}{ii}\,$\lambda$2798
line is in the centre of the $B$ band.

The low-dispersion spectra (in the observer frame) for all objects from 
Table\,\ref{QSO_list} are shown in Fig.\,\ref{allspectra}. The spectra
of several objects show absorption features, most notably
the BAL J171308.8+420811 (no.74) and the luminous high-redshift QSO
no.92. In other respects, the spectra of the VPM 
QSOs are not unusual, in agreement with what was found in Paper~2.

\begin{figure}[hbpt]
\resizebox{8.8cm}{11.8cm}{\includegraphics{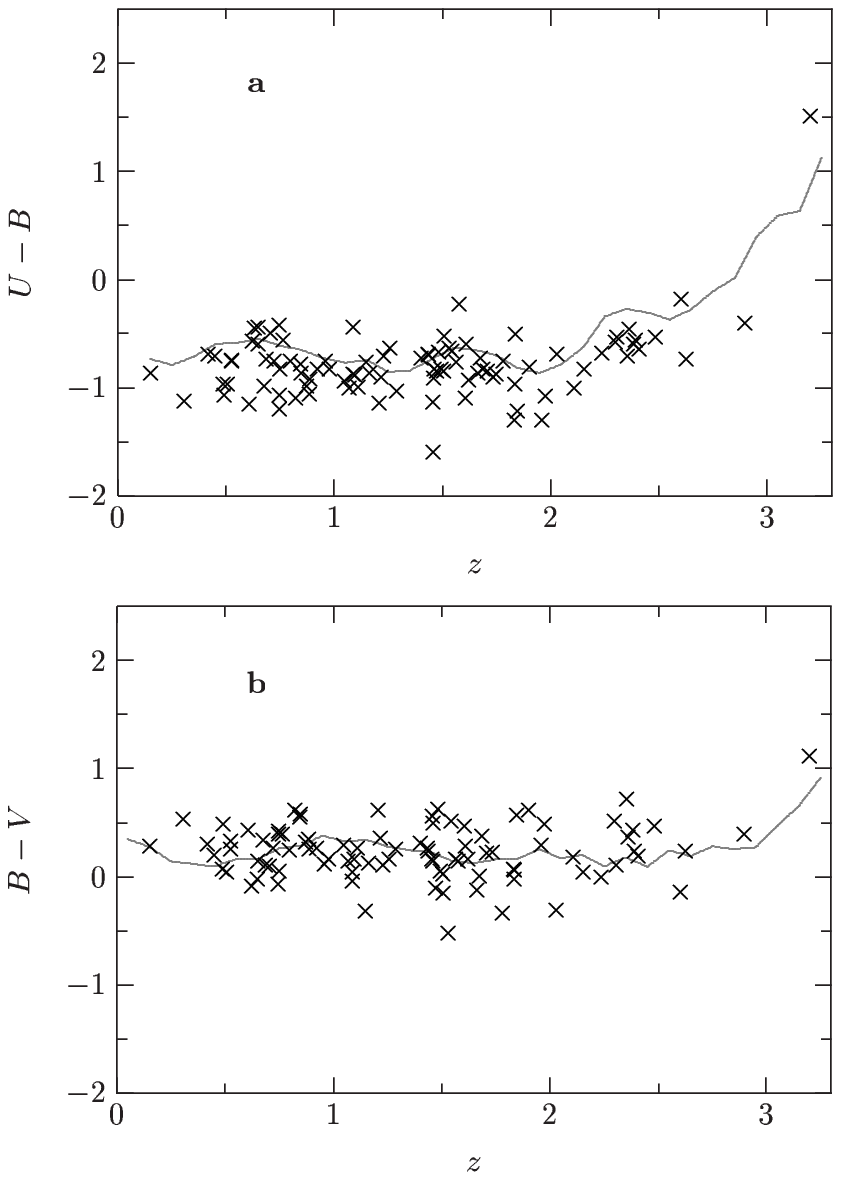}}
\caption{\label{color_z}
Colour indices \textit{U--B} ({\bf a}) and \textit{B--V} ({\bf b}) 
of the identified QSOs ($\times$) as a function of the
redshift $z$. For the sake of comparison, the mean relations for the 
QSOs from V\'eron-Cetty \& V\'eron(\cite{Ver01}) 
are plotted. 
}
\end{figure}

In Paper~1, we presented the colour-colour diagram of the QSO candidates
which shows a broad scatter of the colour indices and a 
substantial fraction of red QSO candidates. In Paper~2 (Fig.4), we have 
shown that all of the high-priority candidates with unusually
red colours proved to be foreground stars and, moreover, that
all previously established VPM QSOs would have been detected by a 
classical two-colour search as well. This result remains true 
for the extended VPM sample, too. 
There is no object among the  QSOs in Table\,\ref{QSO_list} with an 
unusual position either
on the colour-colour diagram (Fig.\,\ref{colour_colour})  or on 
the colour-redshift diagrams (Fig.\,\ref{color_z}). A fraction of 
QSOs with redder colours among the VPM-selected objects is
expected if
(1.) there exists a substantial population of 
QSOs whose continuum is redder than classical colour-selected AGNs
(e.\,g., Webster et al. \cite{Web96}; Kim \& Elvis \cite{Kim99};
Risaliti et al. \cite{Ris01}; Maiolino et al. \cite{Mai01}),
(2.) there is no strict dichotomy of the colour indices, i.e.
there also exists a fraction of QSOs that are mildly redder than
colour-selected ones and not completely dark in the optical
(White et al. \cite{Whi00};
Menou et al. \cite{Men01}), and 
(3.) abnormal red QSOs are in general not much less variable than
colour-selected QSOs.
In this context we note that one explanation of the BAL phenomenon
is related to dust-enshrouded QSOs (Voit et al. \cite{Voi93};
Becker et al. \cite {Bec97}; Egami \cite{Ega99}) and that the
three VPM QSOs with strongest signs of absorption (no. 3, 74, and 92)
all have both strong overall variability and high long-term
variability indices. However, the colour indices of these three
objects do not significantly deviate from the mean colour-$z$ relation
(Fig.\,\ref{color_z}).  In Paper~2, we have estimated that the
fraction of red QSOs must be less than 5\% for $B\le 19.7$.
The  re-evaluation on the basis of the extended QSO sample gives
an upper limit of 3\% for $B\le 19.8$.

\begin{figure}[hbpt]
\resizebox{8.8cm}{6.6cm}{\includegraphics{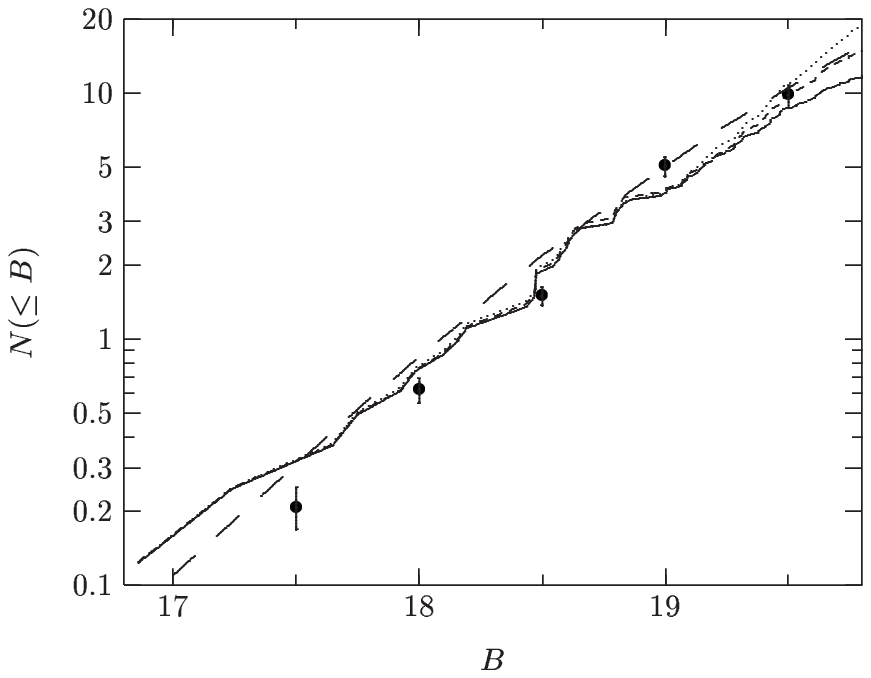}}
\caption{\label{fl_dichte}
Cumulative QSO surface density $N (\le B)$, i.\,e. number of QSOs 
brighter than a given magnitude $B$ per square degree where
$B$ has been corrected for an interstellar
extinction of $A_B=0.08\,$mag. 
{\bf Solid curve:} all VPM QSOs, 
{\bf short-dashed curve:} high-priority VPM QSOs, corrected for 
incompleteness due to the $I_\sigma\ge2$ criterion (Fig.\,\ref{completeness}),
{\bf dotted curve:} high-priority VPM QSOs, corrected for incompleteness due to
the combined $I_\sigma\ge2$ and $I_\Delta\ge2$ criteria assuming periodicities 
$T\approx3$\,yr (see text + Fig.\,\ref{completeness}).
Bullets with error bars: integral surface densities from 
Hartwick \& Schade (\cite{Har90}).
The long-dashed curve corresponds to an analytical approximation
given by Wisotzki (\cite{Wis98}) that was derived from a composite 
optical QSO sample.
} 
\end{figure}

The VPM sample covers a brightness range of three magnitudes,
from $B=16.8$ to 19.9. The cumulative surface density, i.\,e. the
number counts $N(<B)$ per solid angle for the extended VPM QSO
sample is shown in  Fig.\,\ref{fl_dichte}.  The $N(<B)$
roughly follows a power law with an increase of the slope from
about 0.6 at brighter magnitudes to about 0.75 for $B>18$, but
our sample is probably too small for a discussion of the
fine structure of the $N(<B)$ relation.
The main importance of this diagram is to illustrate the
completeness of the survey by the comparison with the number
counts from other surveys. For comparison, Fig.\,\ref{fl_dichte}
shows also the QSO surface density derived by
Hartwick \& Schade ({\cite{Har90}) from a combination of
several optical surveys available at that time.
A mean relation given by Wisotzki ({\cite{Wis98}) is also shown.
The latter
counts have been derived from a combination of more recent
QSO surveys, including the LBQS, the AAT, and the Hamburg/ESO
survey and are given in the form of a polynomial
approximation for $\log\,N(B_J)$. For the transformation
into the Johnson $B$ system  we adopted $B = B_J +0.1$ for QSOs.
There is a good agreement of our number counts with the
comparison relations, except for the fact that our surface
densities at brighter magnitudes are larger than those
given by Hartwick \& Schade, as was mentioned already in Paper~2.
At $B \ga 19.5$, the VPM sample is expected to be substantially
incomplete; the incompleteness corrections in Fig.\,\ref{fl_dichte}
were derived from Fig.\,\ref{completeness}.

The present VPM search is not suited to detect very close pairs of
QSOs since objects have to be separated by at least $8''$ to be included in our
sample as different objects (Paper~1). Despite this restriction, which 
has of course 
no significant influence on the efficiency of the search method,
we looked for QSO pairs within a search radius of $2'$. Two
pairs were found, however with large separations of $91''$
(no.5,84) and 106$''$ (no. 49,98), respectively. In both cases,
the separations in redshifts are very large.

%
\section{Conclusions}

In our previous work (Papers~1 and 2), we have described a
combined VPM QSO search on 162 digitised Schmidt plates of the 
M\,92 field taken with
the Tautenburg Schmidt in the $B$ band between 1960 and 1997.
We have confirmed that this method provides an efficient technique 
for finding AGNs. The efficiency and the completeness of the VPM survey 
primarily depends on the photometric accuracy. The previous QSO sample suffered
from substantial incompleteness at $B>19$. Therefore, we have revised 
the reduction of all $B$ plates on the basis 
of the SExtractor package (Bertin \& Arnouts \cite{Ber96}).  
Both the photometric accuracy and the star-galaxy separation at $B>19$
were substantially improved. The mean photometric accuracy is now 
$\sigma \approx 0.1$\,mag at $B=19$ and $\sigma \approx 0.2$\,mag at $B=20$. 
With the refined variability indices, a number of new QSO candidates 
of medium or high priority were selected. Spectroscopic follow-up 
observations of 84 new candidates revealed  an additional 37 QSOs and 7 
Seyfert\,1s with $B<19.9$.  The total VPM sample in the M\,92 field 
now comprises 
95 QSOs and 14 Seyfert\,1s with $B\le19.9$. For all these 
AGNs long-term lightcurves are available with a baseline of more than 
three decades. Among the 92 high priority QSO candidates with 
$18 \le B \le 19.8$,  we found 70 QSOs and 9 Seyferts\,1s, 
corresponding to a high success rate of 86\%. The completeness of the 
high-priority subsample 
is estimated to be about 87\% for $B\le19$ and 55\% for $B\le19.8$.
A total completeness of 89\% is derived for the sample of all 
QSOs brighter than $B=19.5$ from the comparison with the number 
counts given by  Hartwick \& Schade (\cite{Har90}).
Remarkably, the properties of the VPM QSOs do not significantly 
differ from those of QSOs found by more traditional optical survey
techniques. The deeper VPM sample from the present study confirms the 
result from Paper~2: the spectra and the optical broad band colour 
indices do not provide any hints on a substantial population of abnormal 
red QSOs up to the survey limit.

%

%
\begin{acknowledgements}
%
This research is based on observations made with the
2.2m telescope of the German-Spanish Astronomical Centre, Calar Alto, Spain.
We acknowledge financial support from the
\emph{Deut\-sche For\-schungs\-ge\-mein\-schaft} under
grants Me1350/8-1 and Me1350/13-1. J.B. acklowledges financial
support from the Th\"uringer Ministerium f\"ur Wissenschaft,
Forschung und Kunst.
We are grateful to L. Wisotzki for sending unpublished material.
This research has made use of the NASA/IPAC Extragalactic
Database (NED) which is operated by the Jet
Propulsion Laboratory, California Institute of Technology, under
contract with the National Aeronautics and Space Administration.

\end{acknowledgements}



\begin{thebibliography}{}

   \bibitem[1997]{Bec97} Becker, R.H., Gregg, M.D., Hook, I.M., et al.
   1997, ApJ 479, L93

   \bibitem[1996]{Ber96} Bertin, E., \& Arnouts, S. 1996, A\&AS 117, 393

   \bibitem[1990]{Boy90} Boyle, B.J., Fong, R., Shanks, T., et al. 1990, MNRAS    243, 1

   \bibitem[2001]{Bru01} Brunzendorf, J., \& Meusinger, H. 2001, A\&A 373, 38
   (Paper~1)

   \bibitem[1998]{Con98} Conti, A., Kennefick, J.D., Martini, P., et al. 1998,
   AJ 117, 645

   \bibitem[1996]{Cristianietal1996} Cristiani, S., Trentini, S., LaFranca, F.,
    et al. 1996, A\&A 306, 395

   \bibitem[2001]{Cro01} Croom, S.M., Smith R.J., Boyle, B.J., et al. 2001, 
   MNRAS, 322, L29

   \bibitem[2000]{Dah00} Dahlem, M., \& Thiering, I. 2000, PASP 112, 148

   \bibitem[1999]{Ega99} Egami, E. 1999, in IAU Symp. 196, Galaxy 
   Interactions at Low and High Redshift,  ed. J.E. Barnes \& D.B. Sanders
   (Dordrecht: Kluwer), 475

   \bibitem[1998]{Gol98} Goldschmidt, P. \& Miller, L. 1998, MNRAS 293, 107
   
   \bibitem[2002]{Gre02} Gregg, M. D., Lacy, M., White, R., et al.
    2002, ApJ 564, 133

   \bibitem[1990]{Har90} Hartwick, F.D.A., \& Schade, D., 1990, 
   ARA\&A 28, 437

   \bibitem[1998]{Has98} Hasinger, G., Burg, R., Giacconi, R., et al. 1998, 
   A\&A 329, 482

   \bibitem[1995]{Hew95} Hewett, P.C., Foltz, C.B., \& Chaffee, F.H. 1995, AJ 109, 14   98

   \bibitem[1994]{Hoo94} Hook, I.M., McMahon, R.G., Boyle, B.J., et al.
   1994, MNRAS 268, 305

   \bibitem[1999]{Kim99} Kim, D.-W., \& Elvis, M. 1999, ApJ 516, 9

   \bibitem[1992]{LaF92} La\,Franca, F., Cristiani, S., \&
   Barbieri, C. 1992, AJ 103, 1062

   \bibitem[2001]{Mai01} Maiolino, R., Salvati, M., Marconi, A., et al. 
   2001, A\&A 375, 25

   \bibitem[2001]{Men01} Menou, K., VandenBerk, D. E., Ivezi\'c, \v{Z}., et al.
   2001, ApJ 561, 645

   \bibitem[2001]{Meu01} Meusinger, H., \& Brunzendorf, J. 2001, A\&A 374, 878
   (Paper~2)

    \bibitem[2002]{MeuB02} Meusinger, H., \& Brunzendorf, J. 2002, A\&A,
   submitted (Paper~3)

   \bibitem[2002]{Meu02} Meusinger, H., Brunzendorf, J., Scholz, R.-D., et al.
   2002, in ASP Conf. Ser., AGN Surveys, Proceedings of IAU Colloquium 184,
   ed. R.F. Green, E.Ye. Khachikian, \& D.B. Sanders, in press

   \bibitem[2001]{Ris01} Risaliti, G., Marconi, A., Maiolino, R., et al. 
   2001, A\&A 271, 37

   \bibitem[2001]{Sch01} Schneider, D.P., Richards, G.T., Fan, X., et al.
   2001, AJ, in press

   \bibitem[1997]{Sch97} Scholz, R.-D., Meusinger, H., \& Irwin, M.J. 1997,
   A\&A 325, 457

   \bibitem[2001]{Sha01} Sharp, R.G., McMahon, R.G., Irwin, M.J., et al. 
   2001, MNRAS 326, 45

   \bibitem[2001]{Ver01}V\'eron-Cetty M.P., \& V\'eron P. 2001,
   Quasars and Active Galactic Nuclei (10th Ed.), A\&A 374, 92

   \bibitem[1993]{Voi93} Voit, G.M., Weymann, R.J., \& Korista, K.T. 
   1993, ApJ 95, 109

   \bibitem[1996]{Web96} Webster, R.L., Francis, P.J., Peterson, B.A., et al. 
   1996, Nature 375, 469

   \bibitem[2000]{Whi00} White, R.L., Becker, R.H., Gregg, M.D., et al. 2000, 
   ApJS 126, 133

   \bibitem[1998]{Wis98} Wisotzki, L. 1998, Astron. Nachr. 319, 5

   \bibitem[2000]{Wis00} Wisotzki, L., Christlieb, N., Bade, N., et al. 2000, 
   A\&A, 358, 77

\end{thebibliography}
\end{document}